\documentstyle[12pt,twoside,fleqn,espcrc1,psfig]{article}
\newcommand{\AmS}{{\protect\the\textfont2
  A\kern-.1667em\lower.5ex\hbox{M}\kern-.125emS}}

\def\explainedfig#1#2#3#4{\hbox to \textwidth{%
\psfig{figure=#1,width=0.6\textwidth}\hss%
{\vbox{\hsize=0.4\textwidth\advance\hsize-2ex%
    \caption[#3]{}\label{#4}\footnotesize\baselineskip=2.5ex #2\vss}}}}

\title{$K^0$-$\Sigma^+$ Photoproduction with SAPHIR}

\author{C.~Bennhold$^{\rm a}$, S.~Goers$^{\rm b}$, J.~Barth$^{\rm b}$, 
W.~Braun$^{\rm b}$, R.~Burgwinkel$^{\rm b}$, J.~Ernst$^{\rm c}$, 
K.H.~Glander$^{\rm b}$,  J.~Hannappel$^{\rm b}$, N.~J{\"o}pen$^{\rm b}$, 
H.~Kalinowsky$^{\rm c}$, U.~Kirch$^{\rm b}$, F.~Klein$^{\rm b}$, 
F.J.~Klein$^{\rm g}$, E.~Klempt$^{\rm c}$, A.~Kozela$^{\rm e}$, 
R.~Lawall$^{\rm b}$, Zhenping Li$^{\rm h}$, J.~Link$^{\rm c}$, 
T.~Mart$^{\rm a,f}$, D.~Menze$^{\rm b}$, W.~Neuerburg$^{\rm b}$, 
M.~Paganetti$^{\rm b}$, E.~Paul$^{\rm b}$, H.~van Pee$^{\rm c}$, 
R.~Pl{\"o}tzke$^{\rm c}$, M.~Schumacher$^{\rm d}$, W.J.~Schwille$^{\rm b}$, 
F.~Smend$^{\rm d}$, J.~Smyrski$^{\rm e}$, H.-N.~Tran$^{\rm d}$, 
M.Q.~Tran$^{\rm b}$, F.~Wehnes$^{\rm b}$, B.~Wiegers$^{\rm b}$, 
F.W.~Wieland$^{\rm b}$, J.~Wi{\ss}kirchen$^{\rm b}$ \\ ~ \\
$^{\rm a}$Center for Nuclear Studies, Department of Physics, The George
Washington University,\\ ~Washington DC 20052, USA \\
$^{\rm b}$Physikalisches Institut, Universit{\"a}t Bonn, 53115 Bonn, Germany \\
$^{\rm c}$ISKP, Universit{\"a}t Bonn, 53115 Bonn, Germany \\
$^{\rm d}$II. Phys. Institut, Universit{\"a}t G{\"o}ttingen, 
37073 G{\"o}ttingen, Germany \\ $^{\rm e}$Jagellonian University, Cracow, 
Poland\\ $^{\rm f}$Jurusan Fisika, FMIPA, Universitas Indonesia, 
Depok 16424, Indonesia\\ $^{\rm g}$TJNAF, Newport News, VA 23606, USA\\
$^{\rm h}$Department of Physics, University of Peking, Beijing, 100871, 
P.R. China}

\begin{document}

\maketitle

\begin{abstract}
Preliminary results of the analysis of the reaction 
$\gamma + p \rightarrow K^0 + \Sigma^+$ 
are presented. We show the first measurement of the 
differential cross section and much improved data for the total cross section 
than previous data.   The data are compared with model predictions from 
different isobar and quark models that give a good description of 
$\gamma + p \rightarrow K^+ + \Lambda$ and 
$\gamma + p \rightarrow K^+ + \Sigma^0$ data in the same 
energy range.  Results of ChPT describe the data adequately at threshold
while isobar models that include hadronic form factors reproduce the data
at intermediate energies.
\end{abstract}

\section{PHYSICS WITH THE SAPHIR-DETECTOR}
SAPHIR (\underline{S}pectrometer \underline{A}rrangement for
\underline{PH}oton \underline{I}nduced \underline{R}eactions)
\cite{schwille} is a magnetic spectrometer setup at the 3.5 GeV 
electron accelerator ELSA in Bonn and is designed for the detection of
multiparticle final states. The physics 
aim is the investigation 
of photon-induced reactions off protons and deuterons, especially the 
photoproduction of associated strangeness, of the lightest vector 
mesons $\rho$, $\omega$ and $\phi$ and of the pseudoscalar particles $\eta$ 
and $\eta^{'}$. With our experimental setup we are able to measure 
total and differential cross sections as well as hyperon polarizations. 
A goniometer setup for experiments with polarized photons is being tested 
and optimized. First results of $\gamma + p \rightarrow K^+ + \Lambda$ and 
$\gamma + p \rightarrow K^+ + \Sigma^0$ were published in Ref. \cite{lindemann}.

%\section{THE SAPHIR DETECTOR}
\begin{figure}[htb]
\vspace{-2.0cm}
\explainedfig{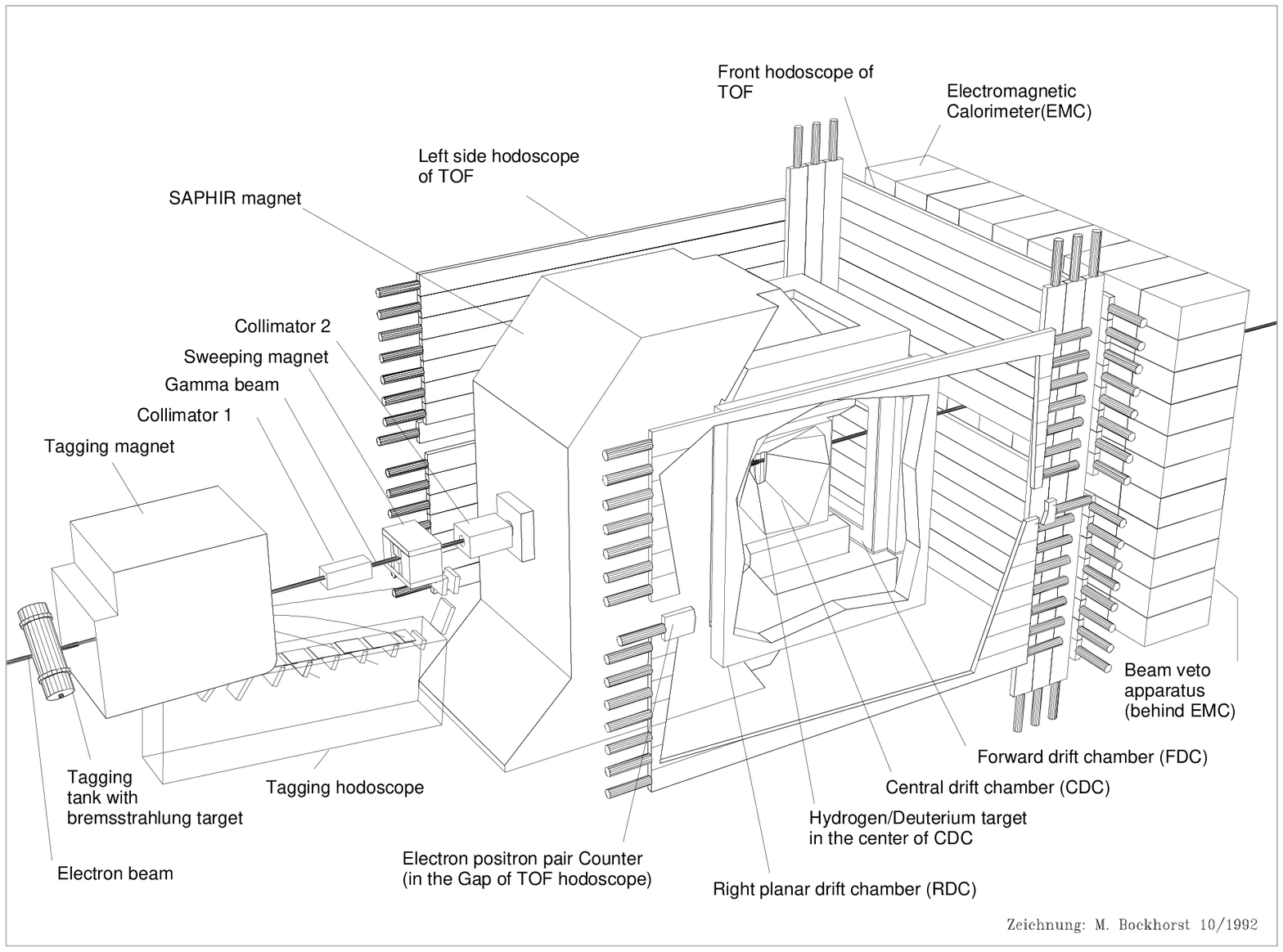}{The SAPHIR detector}{}{saphir_detector}
\end{figure}
The SAPHIR-detector is shown in Fig.~\ref{saphir_detector}. It uses a
tagged photon beam and the data presented here have been taken with a tagging
range of (0.55 - 0.95)$\cdot E$, where E is the energy of the electron beam
from ELSA. The target cylinder, filled with 
liquid hydrogen or deuterium, is located in the gap of a large magnet 
and is surrounded by a drift chamber system, 
consisting of a cylindrical central drift chamber and planar drift 
chambers in forward and sideward directions. They provide the track 
reconstruction and momentum measurement of the charged particles created 
in the target. A system of scintillator hodoscopes serves as trigger 
device and for the determination of particle masses via time-of-flight 
measurements. The electromagnetic calorimeter is currently in a test phase 
and will later be used for the identification of $\gamma$'s coming from 
hadron decays.

\section{DATA REDUCTION AND ANALYSIS}
The data presented here was taken at an electron energy of 1.7 GeV, i.e. 
we cover the photon energy range from threshold ($E_\gamma$ = 1.047 Gev) 
to $E_\gamma$ = 1.6 GeV \cite{goers}. 
Starting from three reconstructed tracks, the first step is to choose a 
subsystem of two tracks with positive and negative charge, with a common
vertex and an invariant mass in the range of the $K^0$ mass. The measured 
3-momenta of the $\pi^+$ and $\pi^-$ (from $K^0$ decay) 
%(xxx-Aenderung-xxx -- andere Formulierung) 
determine the momentum of the $K^0$. Taking the calculated $K^0$ momentum 
and the measured energy of the incident $\gamma$,
the missing mass at the primary vertex must be consistent with $m_{\Sigma^+}$.
The missing $\Sigma^+$ momentum at the primary vertex is 
calculated using a kinematical fit with $\gamma$-energy and $K^0$ momentum as
input. In the next step an iterative vertex fit determines the location of the 
primary vertex and the $\Sigma^+$ decay-vertex simultaneously. The 
%(xxx-Aenderung-xxx -- andere Formulierung) 
identification of the $\Sigma^+$ decay channel
($\Sigma^+ \rightarrow p + \pi^0$ or 
$\Sigma^+ \rightarrow n + \pi^+$) is also obtained by a kinematical fit. The 
last step in the analysis is the separation from background reactions. 

Figure \ref{phys_dist} (A) shows the invariant mass distribution of the $K^0$ 
at the $K^0$ decay vertex, calculated from the measured $\pi^+$ and $\pi^-$ 
tracks; Fig.~\ref{phys_dist} (B) shows the lifetime distribution of the 
$\Sigma^+$ obtained from reconstructed events.
\begin{figure}[htb]
\vspace{-1cm}
\explainedfig{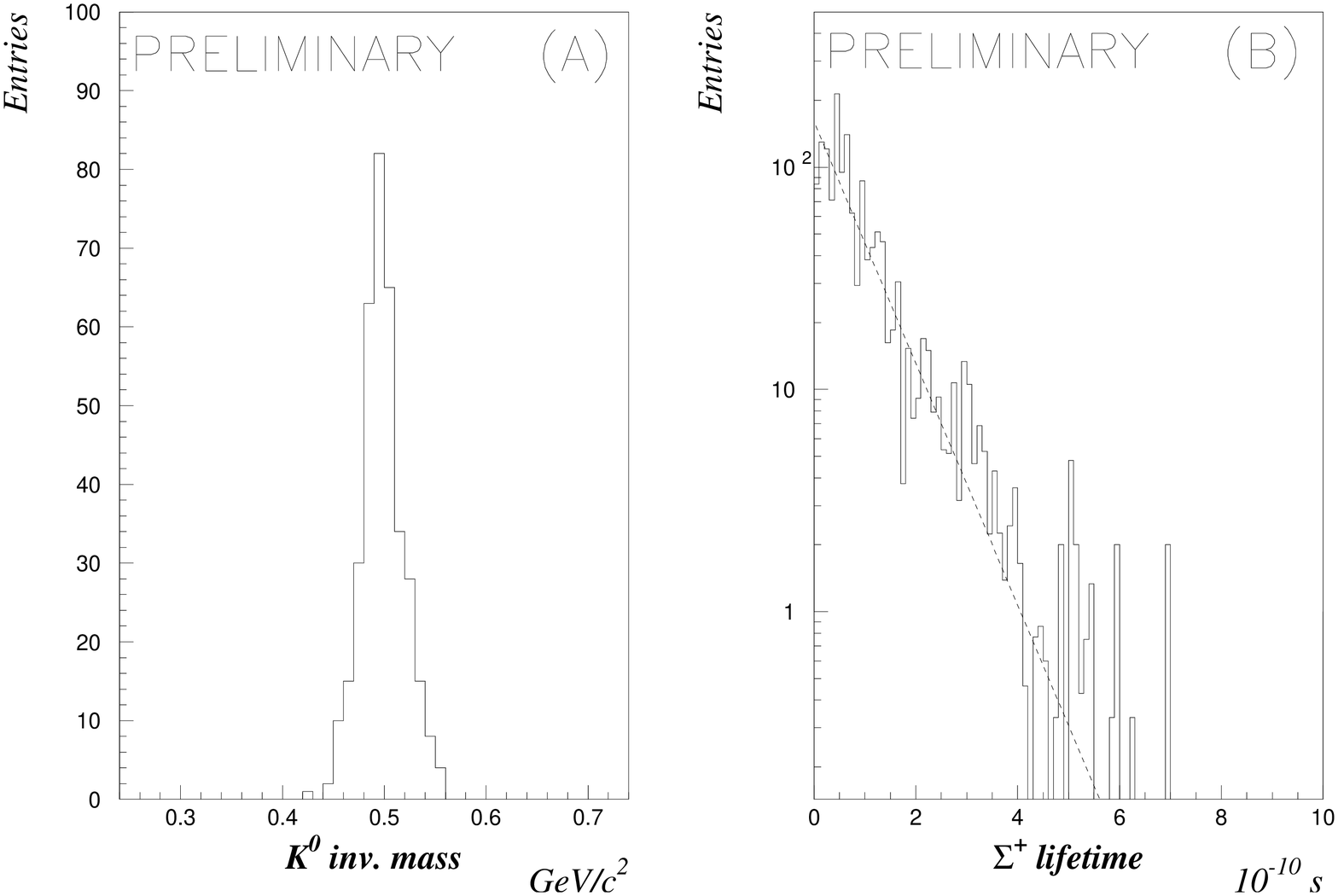}
             {(A) $K^0$ invariant-mass distribution and (B) lifetime of the $\Sigma^+$
               (dashed: PDB-value)}
             {}
             {phys_dist}
% \psfig{file=paw.ps, height=7cm}
% \caption{(A) $K^0$ invariant-mass distribution and (B) lifetime of the $\Sigma^+$
%                (dashed: PDB-value)}
% \label{phys_dist}
\end{figure}

\section{COMPARISON BETWEEN THEORY AND EXPERIMENT}
\begin{figure}[!ht]
\begin{minipage}[htb]{80mm}
{\psfig{figure=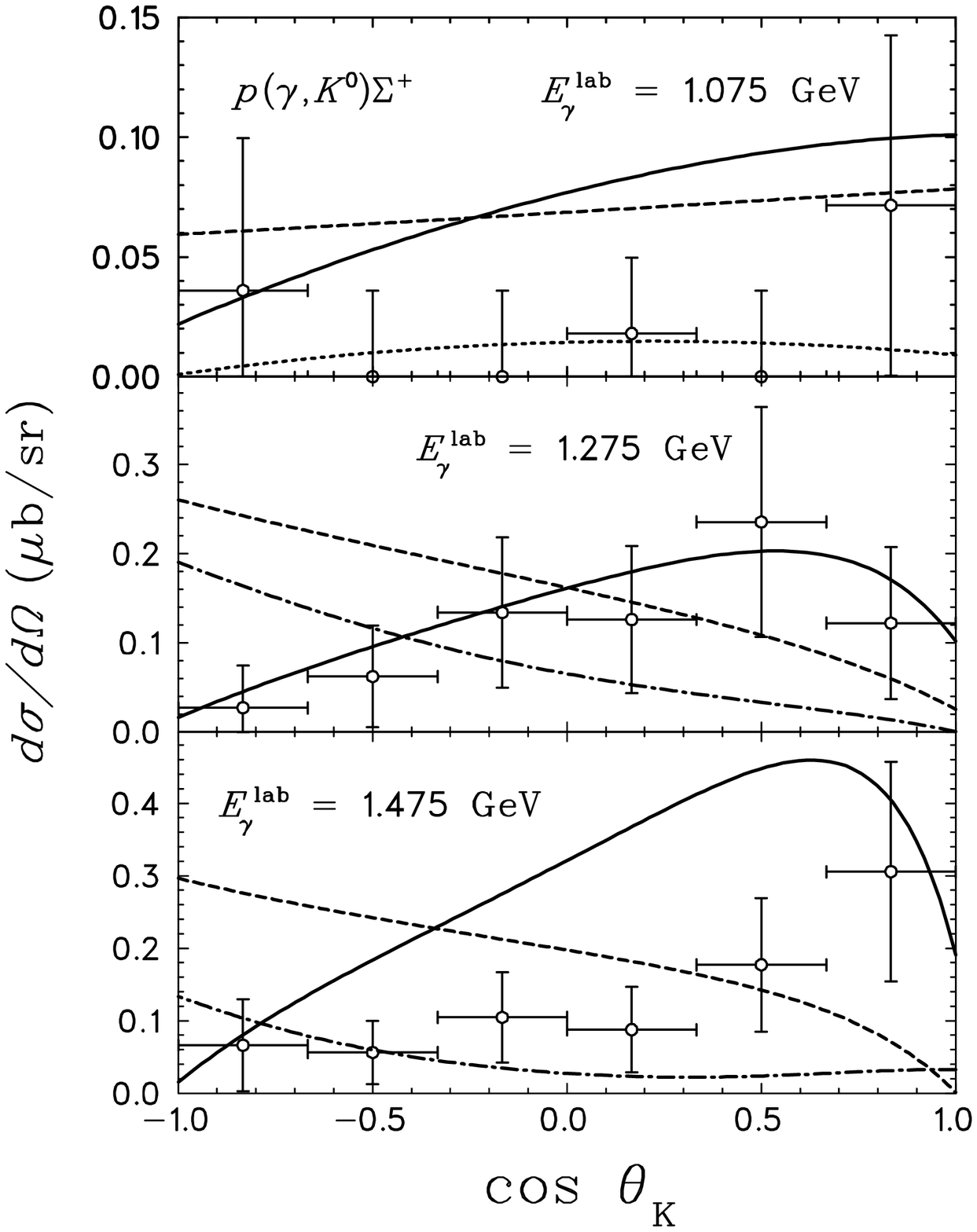,width=75mm}}
\caption{Differential cross sections for $p(\gamma,K^0)\Sigma^+$. The dashed
 (solid) curve shows an isobar model with an overall hadronic form factor 
 \protect\cite{mart} (includes the hadronic form factors according to Ohta 
 \protect\cite{ohta}), while the dash-dotted (dotted) curve shows quark model  
 \protect\cite{li} (ChPT \protect\cite{sven}) calculation.}
\label{fig:diffcs}
\end{minipage}
\hspace{\fill}
\begin{minipage}[t]{70mm}
{\psfig{figure=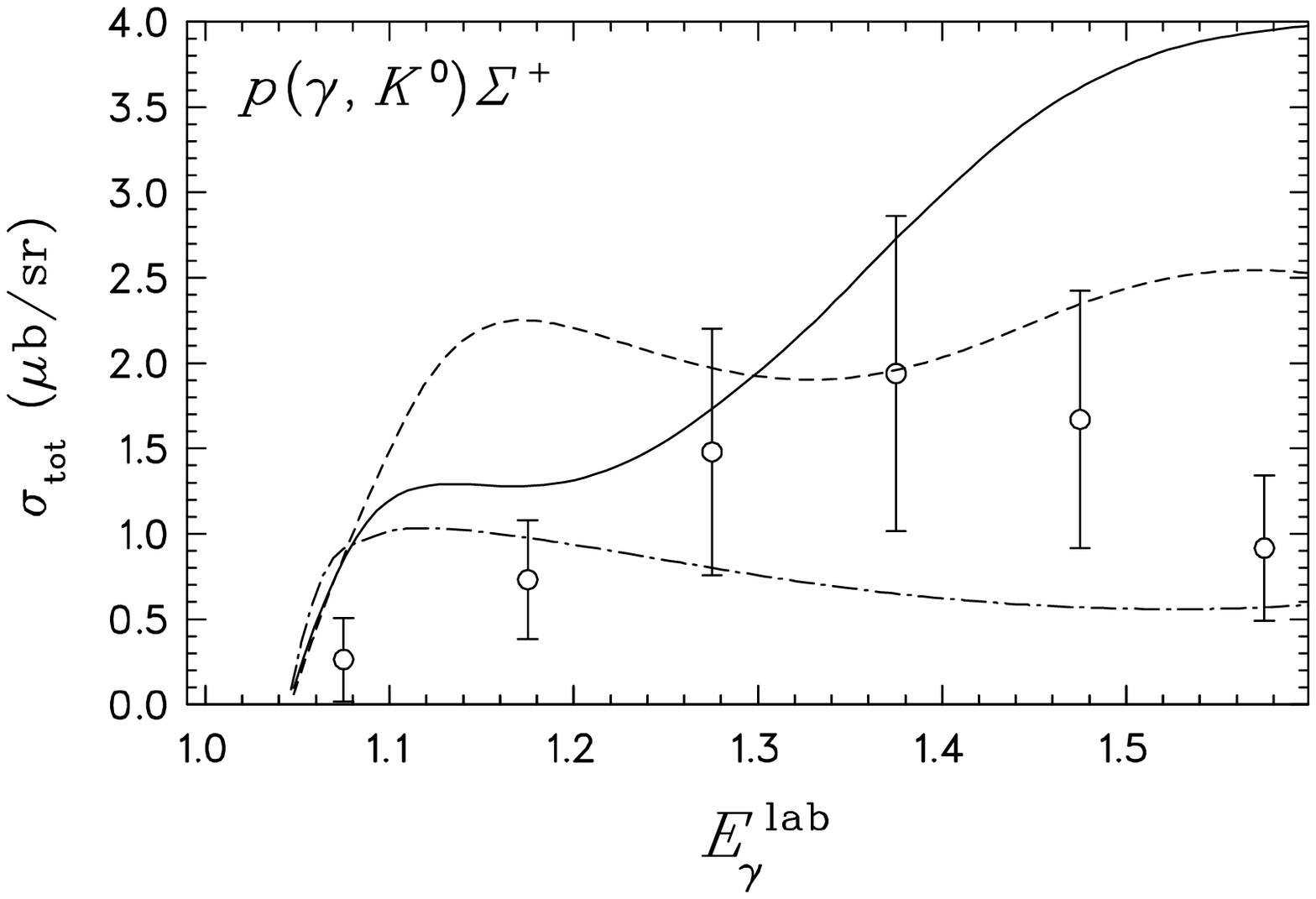,width=75mm}}
\caption{As in Fig. \protect\ref{fig:diffcs}, but for the total cross section.}
\label{fig:totalcs}
\end{minipage}
\end{figure}

The measured differential cross section for the reaction $\gamma + p \rightarrow K^0 
+ \Sigma^+$ as a function of the kaon angle is shown in Fig.~\ref{fig:diffcs}. 
Such data are presented here for the first time, the only
previous data were total cross section data of 
poor statistics \cite{abbhhm}.
Since $K^0 \Sigma^+$ photoproduction is related to $K^+ \Sigma^0$ production
through simple isospin relations both processes should be described by the same
model.  Our data are compared with different 
isobar models \cite{mart,saghai,cotanch}.  In this 
framework, a number of $s$-, $t$- and $u$-channel resonances are 
included at tree level where unknown coupling constants are left as
free parameters to be determined by the data.
Ref. \cite{mart} has for the first time introduced a
hadronic form factor at the $K \Lambda N$ and $K \Sigma N$
vertex. Multiplying the amplitude with an overall monopole form factor
that depends on the momentum transfer $t$ achieves a good
overall fit to all $K^{+}\Lambda$ and $K^{+}\Sigma^{0}$ data
with a cut-off around 700 MeV and allows
 the leading coupling constants
 to assume their SU(3) values.
However, as shown in Fig.~\ref{fig:diffcs},
this fit gives a prediction for $K^{0}\Sigma^{+}$ that is backward-peaked
at higher energies while the data clearly exhibit forward peaking.
Similar results are obtained from the quark model calculation
of Ref.~\cite{li} that has fixed resonance couplings and therefore
far fewer free parameters compared to isobar descriptions. Hadronic form 
factors are provided by the overlap of the quark wave functions. The quark 
model results shown here were performed in the exact SU(3) $\otimes$ O(3) 
limit, thus, quantitative
agreement cannot be expected.  The ChPT result of Ref. \cite{sven} gives an 
adequate description of the data at 1.075 GeV, however, due to its limited 
range of applicability it cannot be compared to data at higher energies.

While multiplying the whole amplitude with an overall form factor
does preserve gauge invariance it represents a theoretically
unjustified (though widely used) approach to include the finite hadron size since
each tree level diagram should depend on a form factor that is
a function of the corresponding off-shell variable.  A naive implementation
of this approach would lead to the violation of gauge invariance,
however, using the procedure outlined in Ref. \cite{ohta} allows
gauge invariance to be restored.  This method leads to a more satisfactory
description of the $(\gamma, K^+)$ data with a cut-off of about 1 GeV for the
$s$- and $u$-channel diagrams and 800 MeV for the $t$-channel resonance 
diagrams. As shown in Fig.~\ref{fig:diffcs}, this model predicts differential 
cross sections for
$K^0 \Sigma^+$ production that are now forward peaked and in good
agreement with the data at intermediate energies, even though
discrepancies appear at higher energies.
The total cross section data as a function of 
the $\gamma$-energy, shown in Fig.~\ref{fig:totalcs}, can 
be reproduced up to 1.4 GeV
in the isobar descriptions while the quark model yields smaller
results and is in better agreement at higher energies.

\end{document}